# Application of Extended Scaling Law to the Surface Tension of Fluids of Wide Range of Molecular Shapes

Mohammad Hadi Ghatee,* Ali Soorghali

(Department of Chemistry, College of Science, Shiraz University, Shiraz 71454, Iran)

E-mail: ghatee@susc.ac.ir

Fax : +98 711 228 6008

___

* Corresponding Author

## Abstract

A linear correlation is presented between the reduced surface tension $\sigma^*$ and reduced temperature $T^*_{extScal}$ by applying the extended scaling law. The correlation is applied quite accurately to 17 atomic, diatomic, and molecular fluid hydrocarbons of wide range of molecular shapes. The reduced surface entropy $S^{s*}_{extScal}$ is introduced, which has a value of 1.000 over the whole liquid range, indicating that the corresponding states principle is followed fully. The correlation for $S^{s*}_{extScal}$ contains a quantity $E^s_\mu$, which is related to the surface energy $E^s$, in the form $E^s_\mu = [\sigma - (1/\mu)T(\partial\sigma/\partial T)]$, where $\sigma$ is the surface tension, $T$ is the absolute temperature, $\mu$ is the critical exponent for surface tension, and the surface entropy $S^s = -(\partial\sigma/\partial T)$. $E^s_\mu$ is different form $E^s$ by an additional $(1/\mu)$ factor coupled to $S^s$. The form of the equation of $E^s_\mu$ is different from the equation of $E^s$, which is a combination of the first and the second laws of thermodynamic for interface. The relation for $S^{s*}_{extScal}$ acts as an intermediate equation to derive a new analytical expression for $E^s_\mu$ in terms of intensive physical and thermodynamic properties of the particular fluid.

Key words: Critical exponent; Extended scaling; Law of corresponding states; Surface energy; Correlation for Surface tension;

# 1. Introduction

The behavior of liquids at the interface, although investigated widely, is still a demanding field of research. The behavior of a liquid-vapor interface is usually characterized by the surface tension, which decreases linearly with temperature close to the freezing point. As the critical point is approached, surface tension decreases non-linearly and becomes zero at the critical point. In this respect, the development of a unified theory that can explain the behavior of surface tension over the whole liquid range is demanding. Empirical and semi-empirical methods are widely available, and are powerful tools for prediction and validation. The law of corresponding states plays an important role in the prediction of surface tension and very often leads to understanding different aspect of the thermodynamic of interfaces.

In a recent letter [1], based on the corresponding states principle and consideration of scaling laws, we have reported a semiempirical linear correlation between reduced surface tension $\sigma^*$ and the reduced temperature $T^*$. The correlation is quite accurate when applied to $C_5$-$C_8$ liquid hydrocarbons from freezing point up to and at the critical point. As a result of these studies, it has been recognized that the expression for surface energy, which brings together surface entropy and surface free energy terms, may include an additional factor of the inverse of critical exponent coupled with the surface entropy term. Thus, the corresponding states principle and scaling laws for surface tension over an extended range of temperature becomes valid if the thermodynamic expression for the surface energy is modified by including its critical exponent, a generic property of fluids under investigation. In general, the rigor of thermodynamics is an important task in surface science because all specific surface effects namely ordering, capillary waves, and double layer influence the surface entropy and its magnitude. Modeling such properties will initiates work to evaluate systems that show critical behavior as a special feature important for industrial application such as food, drug, adhesive, cosmetic, and hygiene industries.

Earlier it has been found that the reduced surface tension [1],

$$\sigma^* = (\sigma/\sigma_f)(T_f/T)^\mu \tag{1}$$

is a linear function of reduced temperature

$$T^* = [(\tau/\tau_f)(T_f/T)]^\mu \tag{2}$$

where $\sigma$ is the surface tension, $T$ is the absolute temperature, and subscript f stands for freezing point. $\tau = (1 - T/T_c)$, with $T_c$ being the critical temperature. $\mu$ is the critical exponent for surface tension. Its theoretical value of 1.26 is universal for the class of liquid hydrocarbons of wide range of molecular shape [2]. The relation between $\sigma^*$ and $T^*$ has been founded semi-empirically. It has been based on the law of corresponding states and phenomenological scaling law, and inspired by the renormalization group theory.

The linear dependence of $\sigma^*$ as function of $T^*$ over the whole liquid range, has led us to suggest an analytical expression for the reduced surface entropy $S^{s*}$ in the form [1],

$$S^{s*} = \frac{d\sigma^*}{dT^*} = E_\mu^{s*} \Big/ \tau^* \qquad (3)$$

where,

$$\tau^* = \left(\tau^{\mu-1} \Big/ \tau_f^\mu\right) \qquad (4)$$

and

$$E_\mu^{s*} = \frac{1}{\alpha}\left(E_\mu^s \Big/ \sigma_f\right), \qquad E_\mu^s = \left[\sigma - \frac{T}{\mu}\left(\frac{\partial\sigma}{\partial T}\right)\right] \qquad (5)$$

where $\alpha$ is a constant close to unity, though it changes slightly with temperature. The value of $S^{s*}$ has been calculated for normal $C_5 - C_8$ liquid hydrocarbons in the temperature range from freezing point up to the critical point (up to $\tau = 10^{-5}$). The average value of $S^{s*}$ has been found to be 0.9990, quite close to unity. $E_\mu^s$ is related to the thermodynamic surface energy $E^s$, except for the factor $(1/\mu)$ coupled with the surface entropy $S^s = -(\partial\sigma/\partial T)$. It has been shown that for normal $C_5 - C_8$ liquid hydrocarbons the $S^{s*}$ is accurately close to unity if the factor $(1/\mu)$ is coupled with the $S^s$ as in the Eq. (5).

In this study, first we present the application of Eq. (3) to a number of monatomic, diatomic, and molecular liquid hydrocarbons of wide range of molecular shape. Second, we pursue and present a more accurate description of the previous expression for $S^{s*}$ by using an accurate analytical relation for the description of experimental surface tension data reported by the application of the extended scaling law. Using the corresponding states principle, a new relation for reduced surface entropy and subsequently surface energy is obtained.

## 2. Derivation of extended reduced surface entropy

Well below the critical point of all known materials, the variation of experimental liquid surface tension can be well approximated by a linear relation of surface tension. Close to the critical point, however, the non-linear vanishing of the surface tension has been noted and investigated as early as van der Waals discussed the continuity of states [3]. The asymptotic behavior of the surface tension close to the critical point can be demonstrated by the scaling law:

$$\sigma = \sigma_\circ \tau^\mu, \qquad \text{for } \tau \to 0 \qquad (6)$$

where $\sigma_\circ$ is a substance dependent constant and $\mu$ is the characteristic universal constant. The critical behavior of fluids has been investigated extensively and based on the scaling law it is found that $\mu = 2\nu$ [5,6], where the critical exponent for the divergence of correlation length $\nu = 0.63$. Based on our definition for the reduced surface tension as $\sigma^* = (\sigma/\sigma_f)(T_f/T)^\mu$ and the application of Eq. (6), $\sigma^*$ is a linear function of $T^*$, and the range of accuracy of this linear behavior has been verified by the application of experimental surface tension data [1].

The extension of the expression for the asymptotic behavior to temperatures close to the freezing point is accounted by the Wegner expansion [7]:

$$\sigma = \sigma_\circ \tau^\mu \left(1 + \sigma_1 \tau^\Delta + \sigma_2 \tau^{2\Delta} + \cdots \right) \qquad (7)$$

where the correction-to-scaling exponent $\Delta = 0.5$. $\sigma_\circ, \sigma_1, \sigma_2$, and $\cdots$ are substance dependent constants, and their values are determined usually by fitting to experimental surface tension data.

Quite obviously, by involving more correction terms in the Eq. (7), a higher accuracy for the description of surface tension over the liquid range is guaranteed. By truncation up to the third term in Eq. (7), now we propose that $\sigma^*$ is a more accurate linear function of

$$T^*_{\text{extScal}} = \alpha_1 \left[ \left(\frac{\tau}{\tau_f}\right)\left(\frac{T_f}{T}\right) \right]^\mu, \quad \alpha_1 = \left( \frac{1 + \sigma_1 \tau^\Delta + \sigma_2 \tau^{2\Delta}}{1 + \sigma_1 \tau_f^\Delta + \sigma_2 \tau_f^{2\Delta}} \right) \qquad (8)$$

where the subscript extScal stands for extended scaling.

For the practical application of Eq. (7) the prediction of surface tension of liquids with acenteric factor $\omega$, the experimental surface tension data of 29 liquids have been used and the fitting parameters were determined [8,9]. These are included Kr, Xe, $O_2$, $N_2$, normal $C_1-C_8$ liquid hydrocarbons, $\text{iso}-C_4$, and 16 refrigerants. Furthermore, from a simple law of corresponding states, it has been argued [8,10] that $\sigma_\circ$ should be proportional to $kT_c(N_A/V_c)^{2/3}$, where $V_c$ is the critical volume, $k$ is the Boltzman constant, and $N_c$ is the Avogadro's number. By considering that the deviation from the law of corresponding states is due to the difference in the molecular shape of liquids under consideration, the following relation, which is similar to an expression from literature [10], has been obtained [8],

$$\sigma = kT_c \left(\frac{N_A}{V_c}\right)^{2/3} (4.35 + 4.14\omega)\tau^\mu \left(1 + 0.19\tau^\Delta - 0.25\tau^{2\Delta}\right) \quad (9)$$

It has been claimed that Eq. (9) predicts surface tension of 29 liquids mentioned above with absolute average deviation of 3.5%, and with 5.9% when applied only to the 16 refrigerants. [8]

By considering the practical applicability of Eq. (9) to describe the asymptotic behavior of surface tension, we now propose that $\sigma^*$ is a linear function of

$$T^*_{extScal} = \alpha_2 \left[\left(\frac{\tau}{\tau_f}\right)\left(\frac{T_f}{T}\right)\right]^\mu, \quad \alpha_2 = \left(\frac{1 + 0.19\tau^\Delta - 0.25\tau^{2\Delta}}{1 + 0.19\tau_f^\Delta - 0.25\tau_f^{2\Delta}}\right) \quad (10)$$

Once the most accurate form of the reduced temperature has been resolved [by Eq. (10)], it is possible to derive an analytical accurate form for reduced surface entropy:

$$S^{s*}_{extScal} = \frac{d\sigma^*}{dT^*_{extScal}} = \frac{1}{\sigma_f \theta^*(T)}\left[\sigma - \frac{T}{\mu}\left(\frac{d\sigma}{dT}\right)\right] \quad (11)$$

where,

$$\theta^*(T) = \left(\frac{\tau}{\tau_f}\right)^\mu \left\{\left(\frac{-0.250 + 0.190\tau^{-\Delta} + \tau^{-2\Delta}}{1 + 0.190\tau_f^\Delta - 0.250\tau_f^{2\Delta}}\right) + \left(\frac{T}{T_c}\right)\left(\frac{-0.250 + 0.0950\tau^{-\Delta}}{1.26 + 0.239\tau_f^\Delta - 0.315\tau_f^{2\Delta}}\right)\right\} \quad (12)$$

The substance dependent variable $\theta^*(T)$ depends on physical properties of the substance e.g., $T_f$ and $T_c$, which are highly accurate measures of the intermolecular forces. The complex form of $\theta^*(T)$ is due to application of the rather complex form of extended scaling law given by the Wegner expansion compared to the simple form which only holds in the critical region. The second term in the expression of $\theta^*(T)$ is negative for $\tau \approx \geq 0.14$ and is much smaller than the first term, however, it is responsible for maintaining the required accuracy close to $T_c$.

Inspection of Eq. (12) shows that $\theta^*(T)$ is dimensionless and involves quotients of temperature $T$ and $\tau$. The asterisks indicates reduced form of (a function of) temperature. Therefore, the reduced surface entropy for an extended temperature range derived in this study is

$$S^{s*}_{\text{extScal}} = E^{s*}_\mu / \theta^*(T) \tag{13}$$

where,

$$E^{s*}_\mu = \left(E^s_\mu / \sigma_f\right), \qquad E^s_\mu = \left[\sigma - \frac{T}{\mu}\left(\frac{\partial \sigma}{\partial T}\right)\right] \tag{14}$$

It is claimed here that Eq. (13) is the most accurate universal relation and its accuracy is determined by the accuracy of Eq. (9).

## 3. Results and Discussion

### 3.A. Reduced Surface Entropy $S^{s*}$

The relation given for the surface tension by the scaling law is valid close to the critical temperature only. Practically it is valid in the temperature range corresponding to $\tau = 10^{-4} - 10^{-5}$. Far from $T_c$, however, by applying the extended scaling law, the accuracy would be extended to well below $T_c$.

In the previous study [1], we have presented the correlation for $S^{s*}$ given by Eq. (3) [see Eq. (3) to (5) for $S^{s*}$]. To arrive at the Eq. (3), no correction to scaling and no adjustment for $\alpha$ has been made. Furthermore, to obtain the input surface tension data, we have applied the values of $\sigma_\circ$ and $\sigma_1$ obtained by Grigoryev et al. by using experimental surface tension data [11]. In this work, we calculate $S^{s*}$ for 9 liquids including normal hydrocarbons $C_1$-$C_8$ and $iso-C4$ (, rather than $C_5$-$C_8$ liquid hydrocarbons considered in the previous study). We use experimental surface tension data described by the Eq. (9) as input data.

Primarily we have calculated the surface tension by application of optimal correction terms to scaling [e.g., $\sigma = \sigma_\circ \tau^\mu (1 + \sigma_1 \tau^\Delta)$]. In each case the plots of $S^{s*}$ versus $\tau$ show that $S^{s*}$ is not constant. Hence, the truncation after the first term gives a surface entropy that does not follow the corresponding states principle. To calculate values of surface tension and its temperature dependence, the corresponding $\sigma_\circ$ and $\sigma_1$ values have been taken from table 11 of reference 8. The plots are shown in Figure 1. As it can be seen the corresponding states principle does not hold even with a moderate accuracy. However, the values of $S^{s*}$ for $C_5$-$C_8$ hydrocarbons, reported in the previous study too [1], are almost constant resulting an average value of 0.9990 for $S^{s*}$.

The same plots, but now including two correction terms [e.g., $\sigma = \sigma_\circ \tau^\mu \left(1 + \sigma_1 \tau^\Delta + \sigma_2 \tau^{2\Delta}\right)$], show non-constant $S^{s*}$ values and show no correlation even with moderate accuracy for normal $C_5$-$C_8$ liquid hydrocarbons. Again, the values of $\sigma_\circ$, $\sigma_1$, and $\sigma_2$ were taken from table 12 of reference 8. Although the application of two correction terms gave some improvement, but the law of corresponding states did not hold.

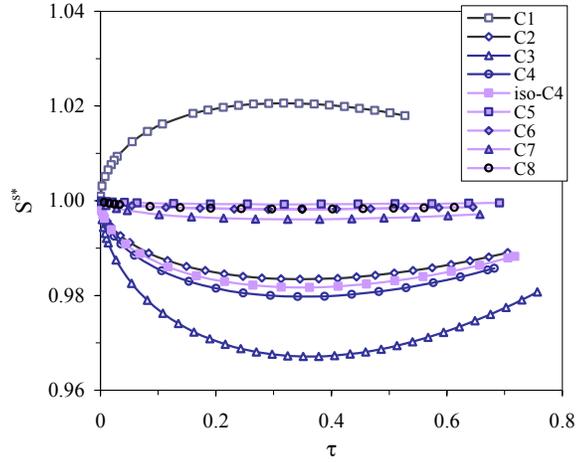

Figure 1. $S^{s*}$ [Eq. (3)] versus $\tau$ for liquid hydrocarbons, according to surface tension correlations including one term correction to the scaling [8]. The lines are trend lines.

Finally, numerically improved correlations for the description of experimental surface tension data are tested to arrive at a rather good correlation for $S^{s*}$ [Eq. (3)] versus $\tau$. In this regards, we have tested two more correlations for the surface tension: the first one is the Eq. (11) of the reference 8, which includes critical parameters as well as acentric factor. The second one is the same except for a numerical improvement, e.g., Eq. (9) above, which has been achieved by fitting the correlation in the surface tension data including additional data covering $O_2$, Xe and Kr gases. The first correlation for surface tension again shows non-constant $S^{s*}$ values and shows no accurate correlation (the plots are not shown). Using the second one, the Eq. (9), the resulted $S^{s*}$ are plotted in Figures 2. In this case 17 liquids including 7 atomic and diatomic fluids (Ar, Kr, Xe, $N_2$, $O_2$, $F_2$, $Cl_2$, $Br_2$,) and 9 molecular hydrocarbons ($C_1$-$C_8$, iso-$C_4$) are tested. In Figure 2, only the results for 9 hydrocarbons are shown. The values of $V_c$ and $\omega$ are taken from references 8 and 12. It can be seen from the figure that the corresponding states principle is better obeyed, however, the accuracy for the $S^{s*}$ value is becoming worse. This improvement can be attributed to the improvements in the accuracy of the applied correlation (for the input surface tension data). The form defined for the variable $T^*$ is simple in that it has not been well characterized by extended scaling law. (see equations 1 to 3). As long as the corresponding states principle is concerned, this can be the cause of non-constant values of $S^{s*}$ obtained over the whole liquids range.

## 3.B. The Reduced Surface Entropy $S^{s*}_{extScal}$

Considering the results of application of different available correlations for surface tension to investigate the features of the previously derived $S^{s*}$, one could conclude that the correlation of surface tension in terms of critical parameters and acenteric factor is accurate enough to result in a high accuracy as long as the corresponding states principle is concerned. Therefore, in this section we consider only the application of Eq. (9) to the developed form of reduced surface entropy, $S^{s*}_{extScal}$. Note that a complete data analysis have been made to employ the most accurate surface tension data available for the construction of Eq. (9) [8]. We have employed four more fluids ($Ar$, $F_2$, $Cl_2$, and $Br_2$) to have a rather complete list of mon- and di-atomic fluids. The values of $T_f$, $T_c$, and $\omega$ for these fluids were taken from reference 12. The accuracy of these data is high enough to compare with the accuracy of the data employed in reference 8 for the other 13 fluids.

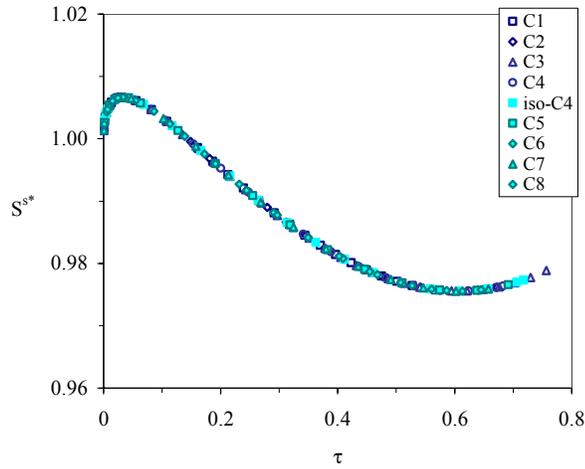

Figure 2. $S^{s*}$ [Eq. (3)] versus $\tau$. The numerically improved correlation for surface tension [Eq. (9)] has been applied.

In this study, we have developed Eq. (13) by using the extended scaling form of $T^*_{extScal}$ given by Eq. (8), and using by the Eq. (9) for surface tension to arrive at $T^*_{extScal}$ given by Eq. (10). This expression for $T^*_{extScal}$ has the advantage of including corrections to scaling (up two terms) over $T^*$. This can be examined by a plot of $\sigma^*$ versus $T^*_{extScal}$ as shown in Figure 3. From a linear fitting of the data in Figure 3, we obtain the slope $(\partial \sigma^* / \partial T^*_{extSacl}) = 1.0$. Thus, $S^{s*}_{extScal}$ exhibit an accurate corresponding states principle because it benefits the accuracy of the extended scaling in addition to the preferential form of the correlation employed for the surface tension. To our opinion this guarantees a high accuracy as long as the corresponding states principle is concerned. The plots of $S^{s*}_{extScal}$ versus $\tau$ are shown in Figure 4.

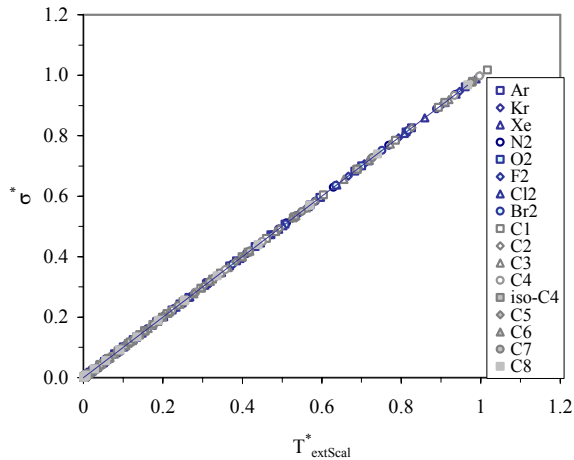

Figure 3. Plots of $\sigma^*$ versus $T^*_{extScal}$ from freezing point to the critical point for 17 liquids.

## 3.C. Surface Energy $E^s_\mu$

The form of $S^{s*}_{extScal}$ involves the same expression for surface energy $E^s_\mu$ as in the previous study [1]. It has been already argued that the form of $E^s_\mu$ could raise the question that the present combined form of first and second laws of thermodynamics for an interface might be different from the results of classical thermodynamic. It involves the factor $(1/\mu)$ coupled with surface entropy. It is of interest that $\mu$ is a constant for all liquids of the universality class and thus it represents the generic behavior of particular universality class. On the other hand, the surface entropy of a number of normal liquid, molten salts, and molten metals are rather close to one another and varies at most by a factor of two. Comparing the values of surface entropies with the corresponding values of surface energies, which vary by 2 to 3 orders of magnitude, it turns out that the surface entropy is a generic property and surface energy is a specific property. Therefore, the character determined for surface energy $E_\mu^s$ involves coupling of two generic properties of the liquid system. This finding deserves more digressing to the fundamentals and in this regards, using the result obtained for 17 liquids shown in Figure 3 and 4,

$$\begin{aligned} E_\mu^s &= \sigma_f \theta^*(T) \\ &= \sigma_f \left(\frac{\tau}{\tau_f}\right)^\mu \left\{ \left(\frac{-0.250 + 0.190\tau^{-\Delta} + \tau^{-2\Delta}}{1 + 0.190\tau_f^\Delta - 0.250\tau_f^{2\Delta}}\right) \right. \\ &\quad \left. + \left(\frac{T}{T_c}\right)\left(\frac{-0.250 + 0.0950\tau^{-\Delta}}{1.26 + 0.239\tau_f^\Delta - 0.315\tau_f^{2\Delta}}\right) \right\} \end{aligned} \qquad (15)$$

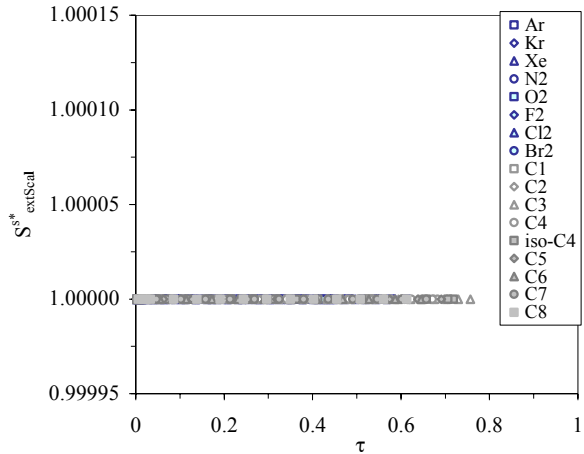

Figure 4. Plots of $S_{extScal}^{s^*}$ versus $\tau$ for liquid hydrocarbons, atomic, and diatomic fluids.

It can be seen that $E_\mu^s$ is a function of $T$ and in addition depends on two generic constants, $\mu$ and $\Delta$. It also depends on fluid specific parameters $\sigma_f$, $T_f$, and $T_c$, which are measures of intermolecular forces. It is worth noting that the final form for $E_\mu^s$ is not involving $\omega$ and $V_c$. This relation for $E_\mu^s$ is reported here for the first time. It involves only intensive thermodynamic and physical properties of the particular liquid. $E_\mu^s$ includes the correction to scaling, however no more fitting is required to arrive at Eqs. (13) and (14).

It seems that the form given by Eq. (15) is uniquely derived by the method presented in this study. First, the expression for $S_{extScal}^{s*}$ is derived, which acts as an intermediate enabling to arrive at the expression for $E_\mu^s$. We noticed that $E_\mu^s$ is derived from $S_{extScal}^{s*}$ without any thermodynamic arguments. One could also derive a relation for $E_\mu^s$ by inserting Eq. (9) appropriately in Eq. (14). In this way, however, the relation resulted for $E_\mu^s$ is not as straightforward as Eq. (15) and would be involved $V_c$ and $\omega$ in addition to the parameters mentioned above.

## 4. Conclusions

We have presented a relation for the reduced surface entropy $S_{extScal}^{s*}$ that is equal to unity over the whole liquid range for monatomic, diatomic molecules, and polyatomic hydrocarbons of the wide range of molecular shapes. Having derived from Eq. (1) and (2), the Eq. (11) for $S_{extScal}^{s*}$ is inspired by the renormalization group theory and phenomenological scaling law. Different relationships have been employed to calculate surface tension required for the calculation of $S_{extScal}^{s*}$. The accuracy of $S_{extScal}^{s*}$ depends highly on the choice of the correlation for the description of experimental surface tension. Using the Wegner extended scaling law, truncated after the second term and treated well by using accurate $\sigma_\circ$, $\sigma_1$, and $\sigma_2$, the $S_{extScal}^{s*}$ has been obtained with high accuracy.

The expression for $S_{extScal}^{s*}$ contains the function $E_\mu^s$. This function is the same as the function for surface energy derived in classical thermodynamics except for a factor of $(1/\mu)$ and is related to the temperature dependence of surface tension $(\partial\sigma/\partial T)$. To derive $E_\mu^s$, Eqs. (1) and (2), which are based on the phenomenological scaling and the principle of corresponding states, has been treated a correlation representing the experimental surface tension data. Variation of $\sigma^*$ versus $T^*$ and or $T_{extScal}^*$ is perfectly linear from $T_f$ to $T_c$. However, by disregarding the critical exponent $\mu$ [in Eqs. (1) and (2)], the variation of $\sigma^*$ versus $T^*$ and or $T_{extScal}^*$, would be rather linear close to $T_f$, but would be non-linear at high $T$'s particularly as $T_c$ is approached. In this case also, the derived surface energy would be $E^s$ (rather than $E_\mu^s$). Noting that the scaling law essentially accounts for the non-linear vanishing of surface tension, a noticeable conclusion is that if one ought to develop a unified theory applicable at low $T's$ as well as high $T's$ close to $T_c$, it (may) requires that the thermodynamic relations be modified and typically it implies that the surface energy could be as $E_\mu^s$ (rather than $E^s$).

The function $S_{extScal}^{s*}$ implies that $E_\mu^{s*} = \theta^*(T)$. Except for the fitting parameters reported in literature, the form of $\theta^*(T)$ has been obtained in an analytical way. $E_\mu^s$ can be determined from knowledge of physical and thermodynamic properties $T_f, T_c, \mu,$ and $\sigma_f$ which all depend on the liquid under consideration. To our knowledge, this relation for surface energy in terms of the mentioned parameters is given here for the first time. However, one should note that $E_\mu^s$ contains the factor $(1/\mu)$, a generic property of the universality classes, which gives a modification to the combination of the first and the second laws of thermodynamic for the interface. This deserves a separate investigation.

**Acknowledgement**

The authors are indebted to the research council of Shiraz University for financial supports. Ali Soorghali is very thankful to the Ministry of Education for granting him leave of absence to perform this research work and the course of study for M.S. degree.

**References**

[1]  M.H. Ghatee, A. Maleki, and H. Ghaed-Sharaf, Langmuir, 19 (2003) 211-213.
[2]  J.V. Sengers, J.M.H. Levelt Sengers, Ann. Rev. Phys. Chem. 37 (1986) 189-222.
[3]  E.A. Guggenheim, J. Chem. Phys. 13 (1946) 253-259.
[4]  C. D. Holcomb, J.A. Zollweg, J. Chem. Phys. 97 (1993) 4797-4907.
[5]  H. Chaar, M.R. Moldover, J.W. Schmidt, J. Chem. Phys. 85 (1986) 418-427.
[6]  M.R. Moldover, Phys. Rev. A 31 (1985) 1022-1033.
[7]  F.J. Wegner, Phys. Rev. B 5 (1972) 4529-4536.
[8]  C. Miqueu, D. Broseta, J. Satherley, B. Mendiboure, J. Lachaise, A. Graciaa, Fluid Phase Equilib., 172 (2000) 169-182.

[9]     C. Miqueu, B. Mendiboure, A. Graciaa, J. Lachaise, Fluid Phase Equilib., 218 (2004) 189-203.

[10]    J.W. Schmidt, E. Carrilo-Nava, M.R. Moldover, Fliud Phase Equilib., 122 (1996) 187-206.

[11]    B.A. Grigoryev, B.V. Memzer, D.S. Kurmov, J.V. Sengers, Int. J. Thermophys. 13 (1992) 453-464.

[12]    R.C. Reid, J.M. Prausnitz, B.E. Poling, The Properties of Gases and Liquids, Fourth Ed. McGraw-Hill, New York 1987.